\documentclass[12pt]{report}
\usepackage[dvips]{graphicx}
\newcommand{\sptwo}{1.4}

\newcommand{\doublespace}{\edef\baselinestretch{\sptwo}\Large\normalsize}

\begin{document}
\doublespace

\begin{center}
{\bf Three-Body Losses in Trapped Bose-Einstein Condensed Gases
}\\
\renewcommand\thefootnote{\fnsymbol{footnote}}
{Yeong E. Kim \footnote{ e-mail:yekim$@$physics.purdue.edu} and
Alexander L. Zubarev\footnote{ e-mail: zubareva$@$physics.purdue.edu}}\\
Purdue Nuclear and Many-Body Theory Group (PNMBTG)\\
Department of Physics, Purdue University\\
West Lafayette, Indiana  47907\\
\end{center}

\begin{quote}
A  time-dependent  Kohn-Sham (KS)-like equation  for
$N$ bosons in a trap is generalized 
 for the case of inelastic collisions. We derive adiabatic equations 
which are used to calculate the nonlinear dynamics of the 
Bose-Einstein condensate
(BEC) and non-mean field corrections due to the 
three-body
 recombination. We find that  the calculated  corrections
 are 
 about 13 times larger for $3D$ trapped dilute bose gases
and about 7 times larger for $1D$ trapped weakly interacting bose gases
 when compared with the corresponding corrections 
for  the ground state energy and for the collective frequencies.

\end{quote}

\vspace{5mm}

\vspace{55 mm}
\noindent

\pagebreak

The newly created Bose-Einstein condensates (BEC) of weakly interacting 
alkali-metal atoms [1] stimulated a large number of theoretical investigations
 [2]. Most of these works are based on the assumption that the properties of
 BEC are well described by the Gross-Pitaevskii (GP) mean-field theory [3].

 According to the theory of Lee, Huang and Yang (LHY) [4] the non-mean-field
 (NMF) correction to
 the GP theory behaves
 like $\sqrt{a^3 n}$, where $n$ is the atomic density and $a$ is the s-wave
 scattering length of interatomic interaction. The gas parameter, $a^3 n$, can
 have a large value when the number of atoms $N$  in the BEC or $a$ is large. 
In recent experiments [5], $N$ was of the order $10^8$ with an 
intermediate value of
the gas parameter, $a^3 n\approx 10^{-3}$.

Recently, it has become possible to tune the atomic scattering length to 
essentially any values, by exploiting the Feshbach resonances (FR) [6-8]. We
 note that the FR do not simply increase the gas parameter [9-13], and we do not consider the FR in this letter.

Theoretical investigations of the NMF corrections to the ground state properties and to the collective frequencies of trapped BEC have already reported in the
 literature [14-18]. 

Inelastic collisions are an important issue in the physics of ultracold gases.
The main goal of this letter is to consider a nonlinear dynamics of the BEC due to the three-body 
recombination. We calculate the corrections to the rate of 
 the three-body recombination due to the NMF  effects.
These corrections are calculated in the large $N$ limit and at zero temperature.

Our starting point is the Kohn-Sham (KS)-like equation [19] for $N$
 interacting bosons in a trap potential $V_{ext}$
$$
i\hbar \frac{\partial \Psi}{\partial t}=-\frac{ \hbar^2}{2 m} \nabla^2 \Psi
+V_{ext} \Psi+\frac{\partial( n \epsilon(n))}{\partial n}\Psi,
\eqno{(1)}
$$
where 
$n(\vec{r},t)=
\mid \Psi(\vec{r},t)\mid^2$  and $\epsilon(n)$ is the ground state energy per
 particle of a uniform system.

In Ref.[19], a rigorous proof is given to show that the KS-like equation
correctly describes properties of the Tonks-Girardeau gas in general
 time-dependent harmonic trap in the large $N$ limit.

The condition for the applicability of Eq.(1) demands that the function $\Psi$
changes slowly on distances of the order of the mean atomic separation $n^{-1/3}
$. Since the characteristic distance of changing $\Psi$ for the trapped gas is
 of the order of the radius of the condensate $R$, one gets the condition
$
N\approx n R^3\gg 1.
$

In order to take into account atoms lost  by inelastic collisions 
(background collisions, dipolar relaxation, three-body recombination, etc.) we
 model the loss by the rate equation
$$
\frac{d N}{d t}=-\int \chi(\vec{r},t) d\vec{r},
\eqno{(2)}
$$
where $\chi(\vec{r},t)=\sum_{l=1} k_l n^l g_l(n)$,  $n^l g_l$ is the local
$l$-particle correlation function and $k_l$ is the rate constant for the
$l$-body atoms loss.
For atoms in BEC, this rate constant is reduced by a factor of $l!$,
 which arises from the coherence properties of condensate
 [20, 21]. 

The generalization of Eq.(1) for the case of inelastic collisions reads
$$
i\hbar \frac{\partial \Psi}{\partial t}=-\frac{ \hbar^2}{2 m} \nabla^2 \Psi
+V_{ext} \Psi+\frac{\partial( n \epsilon(n))}{\partial n}\Psi- i \frac{\hbar}{2} \sum_{l=1} k_l n^{l-1} g_l(n)\Psi.
\eqno{(3)}
$$
We will call Eq.(3) as dissipative KS-like equation. We note that Eq.(3) can be obtained from Eq.(1) using replacement of $\epsilon(n)$ by $\epsilon(n)-i \Gamma/2$, where
$$
\Gamma/2=\sum_{l=1}(\hbar/(2 n)) k_l \int_0^n x^{l-1} g_l(x) dx.
$$

It can be proved that  every solution of the KS-like equation (1) is a stationary point of an action corresponding to the Lagrangian density
$$
\mathcal{L}_0=\frac{i\hbar}{2}(\Psi\frac{\partial\Psi^{\ast}}{\partial t}-
\Psi^{\ast}\frac{\partial\Psi}{\partial t})+\frac{\hbar^2}{2m}\mid \nabla \Psi\mid^2+\epsilon(n)n+V_{ext}n,
\eqno{(4)}
$$
where the asterisk denotes the complex conjugation. Indeed, the substitution of $\mathcal{L}_0$, Eq.(4), into the  following Euler-Lagrange (EL) equations
$$
\everymath={\displaystyle}
\begin{array}{rcl}
\frac{\delta \mathcal{L}_0}{\delta \Psi^{\ast}}=\frac{\partial \mathcal{L}_0}{\partial \Psi^{\ast}}-\frac{\partial}{\partial t}\frac{\partial \mathcal{L}_0}{\partial(\partial \Psi^{\ast}/\partial t)}-
\nabla \frac{\partial \mathcal{L}_0}{\partial(\nabla \Psi^{\ast})}=0,\\ \\
\frac{\delta \mathcal{L}_0}{\delta \Psi}=\frac{\partial \mathcal{L}_0}
{\partial \Psi}-\frac{\partial}{\partial t}\frac{\partial \mathcal{L}_0}
{\partial(\partial \Psi/\partial t)}-
\nabla \frac{\partial \mathcal{L}_0}{\partial(\nabla \Psi)}=0 
\end{array}
\eqno{(5)}
$$
gives Eq.(1) and its complex conjugate equation.

Now for the dissipative KS-like equation (3) we write the corresponding 
Lagrangian $\mathcal{L}$ as a sum of two terms, a conservative one
 $\mathcal{L}_0$, Eq.(4), and nonconservative one $\mathcal{L}^\prime$,
$\mathcal{L}=\mathcal{L}_0+\mathcal{L}^\prime$ [22-24].
Now the EL equations read
$$
\everymath={\displaystyle}
\begin{array}{rcl}
\frac{\partial \mathcal{L}_0}{
\partial \Psi^{\ast}}-\frac{\partial}{\partial t}\frac{\partial \mathcal{L}_0}
{
\partial(\partial \Psi^{\ast}/\partial t)}-
\nabla \frac{\partial \mathcal{L}_0}{\partial(\nabla \Psi^{\ast})}+
\frac{\delta \mathcal{L}^\prime}{\delta \Psi^\ast}=0,\\ \\
\frac{\partial \mathcal{L}_0}
{\partial \Psi}-\frac{\partial}{\partial t}\frac{\partial \mathcal{L}_0}
{\partial(\partial \Psi/\partial t)}-
\nabla \frac{\partial \mathcal{L}_0}{\partial(\nabla \Psi)}+
\frac{\delta \mathcal{L}^\prime}{\delta \Psi}=0
\end{array}
\eqno{(6)}
$$
A comparison Eqs.(6) with Eq.(3) gives
$$
\everymath={\displaystyle}
\begin{array}{rcl}
\frac{\delta \mathcal{L}^\prime}{\delta \Psi^\ast}=
\frac{\partial \mathcal{L}^\prime}{
\partial \Psi^{\ast}}-\frac{\partial}{\partial t}\frac{\partial 
\mathcal{L}^\prime}
{
\partial(\partial \Psi^{\ast}/\partial t)}-
\nabla \frac{\partial \mathcal{L}^\prime}{\partial(\nabla \Psi^{\ast})}=
-\frac{i \hbar}{2}\sum_{l=1} k_l n^{l-1} g_l(n)\Psi,\\ \\
\frac{\delta \mathcal{L}^\prime}{\delta \Psi}=
\frac{\partial \mathcal{L}^\prime}
{\partial \Psi}-\frac{\partial}{\partial t}\frac{\partial \mathcal{L}^\prime}
{\partial(\partial \Psi/\partial t)}-
\nabla \frac{\partial \mathcal{L}^\prime}{\partial(\nabla \Psi)}=
\frac{i \hbar}{2}\sum_{l=1} k_l n^{l-1} g_l(n)\Psi^\ast.
\end{array}
\eqno{(7)}
$$
We are now ready  to rewrite the Hamilton principle $\delta \int dt\int d\vec{r} (\mathcal{L}_0+\mathcal{L}^\prime)=0$ as
$$
\delta \int dtL_0+\frac{i \hbar}{2} \sum_{l=1}k_l\int dt \int d \vec{r}
n^{l-1}g_l(n)(\Psi^\ast \delta \Psi-\Psi \delta \Psi^\ast)=0,
\eqno{(8)}
$$
where $L_0=\int \mathcal{L}_0 d \vec{r}$.
The variational formulation, Eq.(8), is an extension of the standard variational formulation for the situation where a Lagrangian corresponding to the original equations can not be found or does not exist [22-24].

For the remainder of this letter, we will focus and specialize solely on the
 three-body recombination.  A suitable trial function can be taken as $\Psi(\vec{r},t)=
\exp[-(i/\hbar)\phi(t)]\Phi(\vec{r},t)$, where both $\phi$ and $\Phi$ are real functions. With this ansatz, the Hamiltonian principle , Eq.(8), gives 
the following variational equations 
$$
\frac{d N}{d t}=-k_3 \int \Phi^6(\vec{r},t) g_3(\Phi^2(\vec{r},t)) d \vec{r},
\eqno{(9)}
$$
and
$$
-\frac{ \hbar^2}{2 m} \nabla^2 \Phi
+V_{ext} \Phi+\frac{\partial( n \epsilon(n))}{\partial n}\Phi=\mu(t) \Phi,
\eqno{(10)}
$$
where $N(t)=\int \Phi^2(\vec{r},t)d \vec{r}$, and $\mu(t)=d\phi/dt$.

The condition of the validity of the  adiabatic equations (9) and (10) is
 $dN/dt<\omega_{\nu}N$, where $\omega_\nu$ is a frequency of elementary
 excitation.

The ground state energy per particle, $\epsilon(n)$, in the low-density regime
can be calculated using an  expansion in power of $\sqrt{n a^3}$
$$
\epsilon(n)=\frac{2 \pi \hbar^2}{m} an[1+\frac{128}{15 \sqrt{\pi}}
(n a^3)^{1/2}+8 (\frac{4 \pi}{3}-\sqrt{3}) na^3 [\ln(n a^3)+C]+...].
\eqno{(11)}
$$
The first term corresponding to the mean field prediction was first calculated
 by Lenz
 [25]. The coefficient of the $(na^3)^{3/2}$ term (the second term) was first 
calculated by Lee, Huang, and Yang [4], while the coefficient of the last term 
was first obtained by Wu [26]. The constant $C$ after the logarithm term  was
 considered
 in Ref.[27].

The expansion (11) is asymptotic, and it was shown in Ref.[28] that the
Lee-Huang-Yang (LHY) correction (second term in Eq.(11)) represents a
significant improvement on the mean field prediction up to $a^3n\approx 10^{-2}$
,
 but the logarithmic correction already is wrong  at $na^3\approx 10^{-3}$.
In Refs.[16-18], the LHY  expansion (first two terms
in the expansion (11)) has been used to study effects beyond the mean field
 approximation.

We do not consider the logarithmic term in the expansion (11), and rewrite
Eq.(10) in the limit of large $N$ as
$$
\frac{m}{4 \pi a \hbar^2}(\mu-V_{ext})=n(1+\frac{32}{3\sqrt{\pi}} \sqrt{na^3}),
\eqno{(12)}
$$
where $n=\Phi^2(\vec{r},t)$.

At  densities ($na^3\le 10^{-3}$) Eq.(12) can be solved by
 iteration
$$
n=\frac{m}{4 \pi a \hbar^2}(\mu-V_{ext})-\frac{4 m^{3/2}}{3 \pi^2 \hbar^3}(\mu-V_{ext})^{3/2}+\frac{32 a m^2}{3 \pi^3 \hbar^4}(\mu-V_{ext})^2-
\frac{896 a^2 m^{5/2}}{9 \pi^4 \hbar^5}(\mu-V_{ext})^{5/2}+...,
\eqno{(13)}
$$
where
$$
\mu=\mu_{TF}(1+\frac{1}{2} \frac{a m^{1/2} \mu^{1/2}_{TF}}{\hbar}-
(\frac{1024}{105 \pi^2}-\frac{9}{16})\frac{a^2m\mu_{TF}}{\hbar^2}+
(\frac{25}{32}-\frac{22}{21 \pi^2})\frac{a^3 m^{3/2} \mu^{3/2}_{TF}}{\hbar^3}-...),
\eqno{(14)}
$$
and the Thomas-Fermi (TF) approximation is simply obtained by keeping only the first
 term in the right side of Eq.(13), $n_{TF}=m(\mu_{TF}-V_{ext})/(4 \pi a \hbar^2
)$.

Eq.(13) holds in the region where $n$ is positive and $n=0$ outside this region. We note that the second term in Eq.(14) was considered in Refs.[14,18].

For the harmonic trap potential $V_{ext}(\vec{r})=(m/2) (\omega^2_x x^2+
\omega^2_y y^2+\omega^2_z z^2)$, $\mu_{TF}$ is given by [29] $\mu_{TF}=
(\hbar^2/(2 m a^{12/5}_{ho}) (15 a N)^{2/5}$, where 
$a_{ho}=(\hbar/m\omega_{ho})^{1/2}$ is the oscillator length and $\omega_{ho}=(\omega_x \omega_y \omega_z)^{1/3}$. Eq.(13) in this case becomes
$$
\mu=\mu_{TF}(1+\frac{1}{2}(4 \pi a^3 n_{TF}(0))^{1/2} -
(\frac{1024}{105 \pi^2}-\frac{9}{16})4 \pi a^3 n_{TF}(0)+
(\frac{25}{32}-\frac{22}{21 \pi^2})(4 \pi a^3 n_{TF}(0))^{3/2}-..
.),
\eqno{(15)}
$$
where $n_{TF}(0)=m\mu_{TF}/4\pi a \hbar^2$.

Using the correlation function [21]
$$
g_3(n)=(1+\frac{64}{\sqrt{\pi}} \sqrt{n a^3}+...),
\eqno{(16)}
$$
it can easily be seen that
$$
\int \Phi^6(\vec{r},t) g_3(n)d \vec{r}=\frac{8}{21} n^2_{TF}(0) N
 (1+\frac{3357}{512} (4 \pi a^3 n_{TF}(0))^{1/2}+...)
\eqno{(17)}
$$

Solution of Eqs.(9) and (10) in the mean-field approximation, 
corresponding to the first term in the expansion (16) and (17), reads
$$
\Phi^2(\vec{r},t)=\frac{8}{21} (\frac{a^{6/5} a_{ho}^{24/5}}{\alpha N^{4/5}(0)}
+\frac
{4}{5} k_3 t)^{-1/2}-\frac{m}{4 \pi a \hbar^2} V_{ext},
\eqno{(18)}
$$
and
$$
N(t)=(N^{-4/5}(0)+\frac{4}{5} \frac{\alpha k_3}{a^{6/5} a_{ho}^{24/5}}t)^{-5/4},
\eqno{(19)}
$$
where $\alpha=15^{4/5}/168 \pi^2$,
and for the rate $\tau=\mid d\ln N/d t\mid$ we obtain
$$
\tau=\frac{\alpha k_3/a^{6/5} a_{ho}^{24/5}}
{N^{-4/5}(0)+(4/5) (\alpha k_3/a^{6/5} a_{ho}^{24/5}) t}.
\eqno{(20)}
$$

Analytical results, Eqs.(18-20), predict a strong $\omega_{ho}$ dependence of
 the 3-body recombination in the TF regime, as shown for 
$^{87}Rb$ condensate with $k_3=5.8\times 10^{-30} cm^6/s$ [20] 
in Fig 1. For all cases of Fig.1, the adiabaticity is insured ($\dot{N}/N \leq
4\times 10^{-4} \omega_{ho}$ and $\dot{N}/N\leq 5\times 10^{-2} \omega_{ho}$ for
$\omega_{ho}/2\pi=12.83$Hz and 77.78Hz respectivelly).

A comparison of  Eq.(17) with Eq.(15) shows that  
 the  non-mean-field corrections to the nonlinear
 dynamics of the BEC
 due to the three-body recombination are about 13 times larger than
 corresponding corrections to the ground state and to the collective
 frequencies of trapped BEC. As an example, we present in Fig. 2
the results of numerical
calculation of the non-mean-field corrections to the rate 
$\tau=\mid d \ln N/d t\mid$ for the BEC of $^{87} Rb$ atoms with $N(0)=10^8$.

For a  $1D$ Bose gas interacting via a repulsive $\delta$-function potential,
$ \tilde{g} \delta(x)$,  the Lieb-Liniger (LL) model [30],
$\epsilon(n)$ is given by [30]
$
\epsilon(n)=\frac{\hbar^2}{2 m} n^2 e(\gamma),
$
where $\gamma=m\tilde{g}/(\hbar^2 n)$ and
for small values of $\gamma$, the following expression for $\epsilon(n)$,
$
\epsilon(n)=\frac{\tilde{g}}{2} (n-\frac{4}{3 \pi}\sqrt{\frac{m\tilde{g} n}
{\hbar
^2}}+...)
$
is adequate up to approximately $\gamma=2$ [26].
In this case the large N limit solution of Eq.(10) reads
$$
n=\frac{\mu-V_t}{\tilde{g}}+\frac{1}{\pi}\sqrt{\frac{m\tilde{g}}{\hbar^2}}
(\frac{\mu-V_t}{\tilde{g}})^{1/2}+...,
\eqno{(21)}
$$
where
$$
\mu=\mu_{TF} (1-\frac{1}{4}\sqrt{\gamma(0)}+...),
\eqno{(22)}
$$
with $\gamma(0)=m\tilde{g}/\hbar^2 n(0)$ and  for $V_t=m\omega^2 x^2/2$,
$\mu_{TF}=(3 N \tilde{g}m^{1/2} \omega/2^{5/2})^{2/3}$.
 Using the $1D$ correlation function [31] $g_3(n)=1-6\sqrt{\gamma}/\pi+...$,
we obtain
$$
\int \Phi^6(x,t) g_3(n) dx=\frac{24}{35} n_{TF}^2(0) N (1-\frac{973}{512}\sqrt{\gamma(0)}+...),
\eqno{(23)}
$$
where $n_{TF}(0)=\mu_{TF}/\tilde{g}$.

The last equation (23) shows that for the weak interacting trapped $1D$ bosons the NMF corrections to the three-body recombination dynamics are about 7 times larger than corresponding corrections to the ground state energy.

In conclusion, we have developed a
  time-dependent dissipative Kohn-Sham (KS)-like equation  for
$N$ bosons in a trap
 for the case of inelastic collisions. We derive adiabatic equations
which are used to calculate the nonlinear dynamics of the BEC due to the
three-body
 recombination. The calculated  non-mean-field corrections
 to the three-body recombination dynamics
 are
shown to be about 13 times larger for $3D$ trapped dilute bose gases
and about 7 times larger for $1D$ trapped weak-interacting bose gases
 than corresponding corrections
to the ground state energy and to the collective frequencies.

We thank S. Khlebnikov for his interest and comments.

\pagebreak

\begin{figure}[ht]
\includegraphics{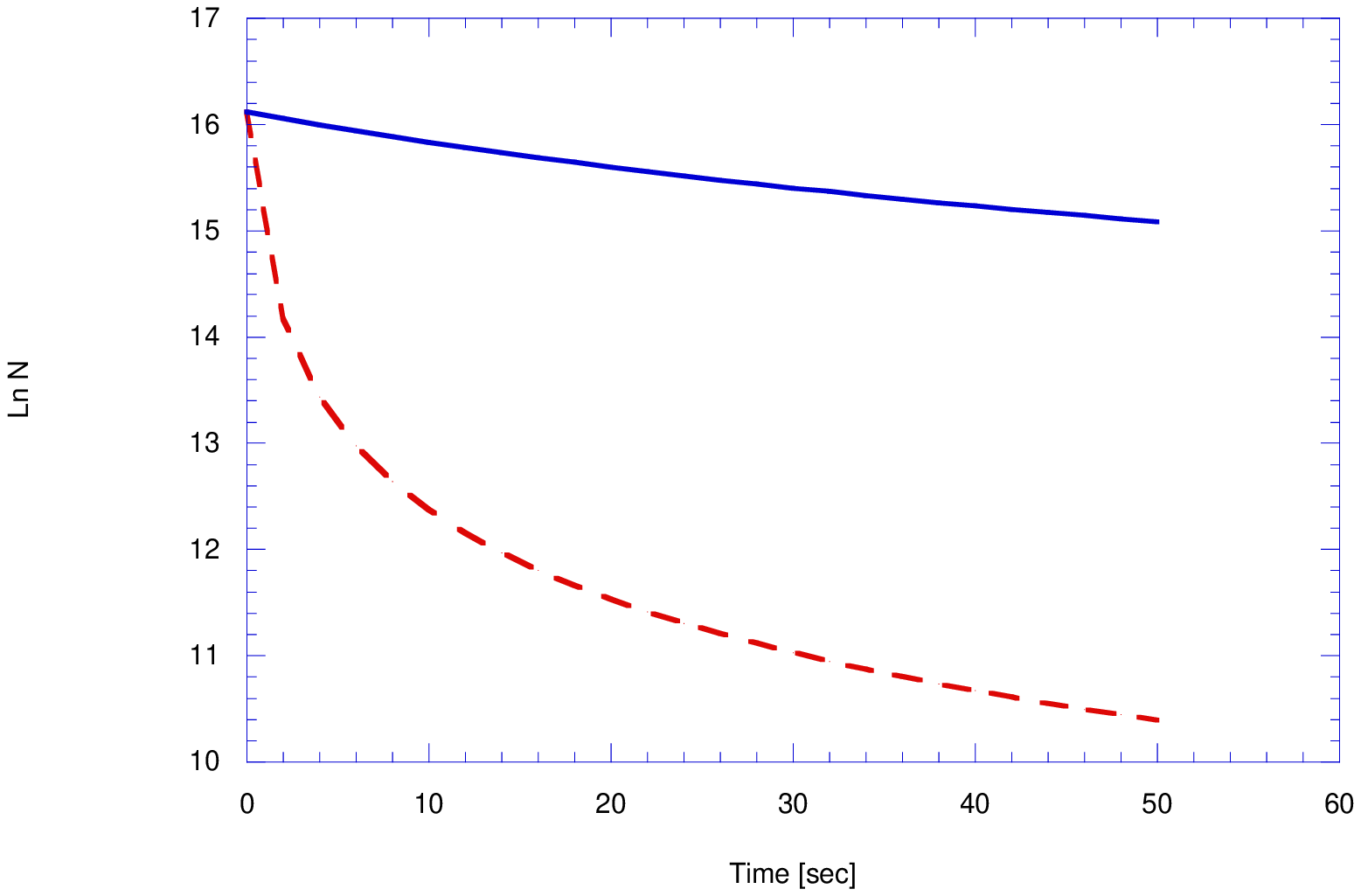}
\end{figure}

\vspace{10mm}
FIG. 1. The natural {\it log} of the number of $^{87} Rb$ atoms of the BEC 
in the trap as a function of time. The loss is due to the three-body 
recombination. Solid line and dashed line represent the theoretical results
 for  traps with geometric average of the oscillator frequencies
 $\omega_{ho}/2 \pi=12.83$Hz  and $\omega_{ho}/2 \pi=77.78$Hz,
 respectively.

\pagebreak

\begin{figure}[ht]
\includegraphics{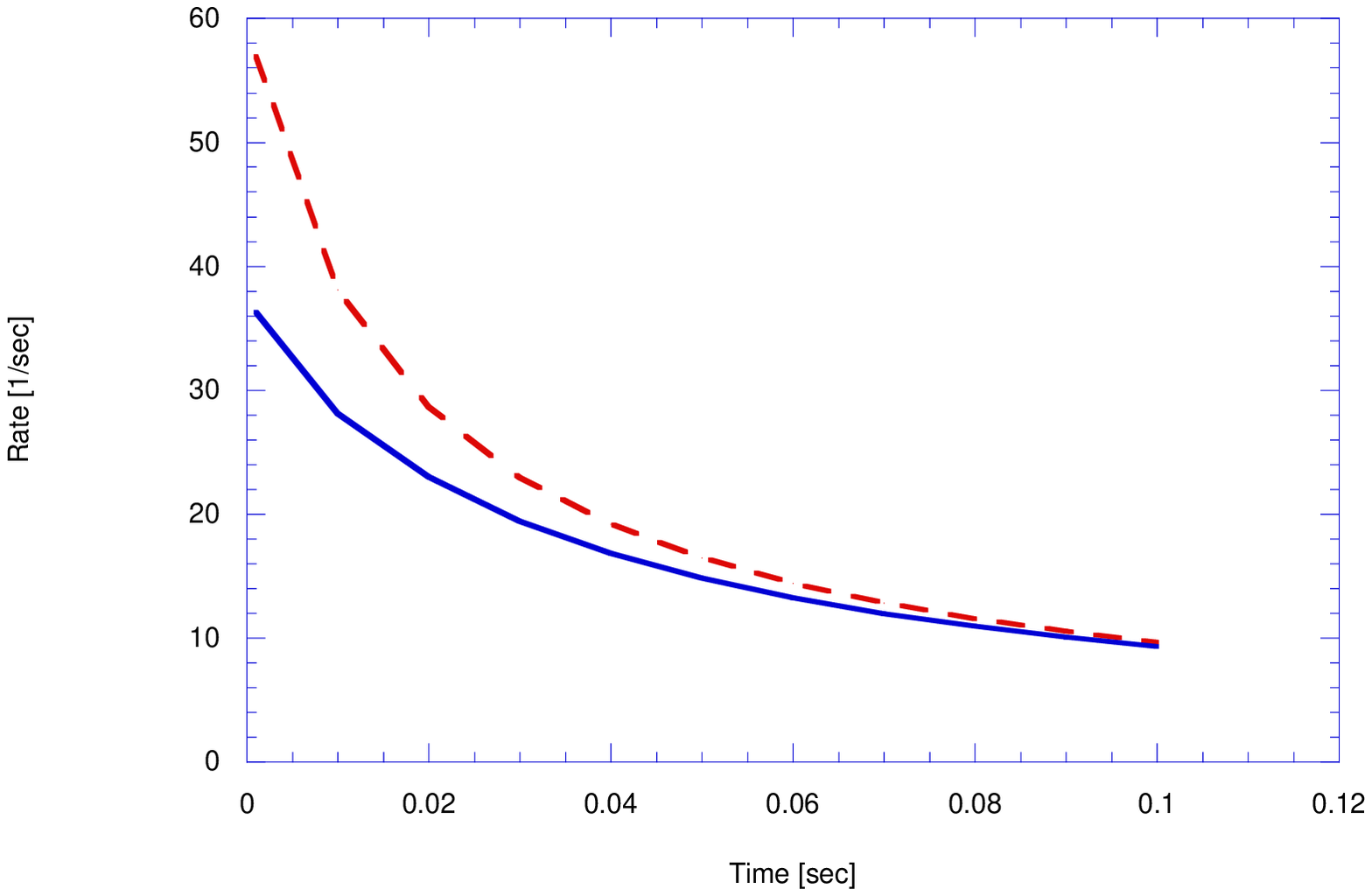}
\end{figure}

\vspace{10mm}

FIG. 2. Rate of the three-body recombination, $\tau=\mid \frac{d lnN}{dt} \mid$, as a function of time for the BEC of $^{87}Rb$ atoms for the initial condition
 $N(t=0)=10^8$. The asymmetry parameter of the trap is $\lambda=\sqrt{8}$ and $\omega_z/(2 \pi)=220$Hz. The solid line corresponds
to the mean-field approximation  and the dashed line 
shows the results of inclusion the  non mean-field corrections. 

\pagebreak

{\bf References}
\vspace{8pt}

\noindent
1. {\it Bose-Einstein Condensation in Atomic Gases}, Proceedings of the
International
 School of Physics ``Enrico Fermi", edited by M. Inquscio, S. Stringari,
 and
 C.E. Wieman (IOS Press, Amsterdam, 1999);http://amo.phy.gasou.edu/bec.html;
http://jilawww.colorado.edu/bec/
 and references therein.

\noindent
2.  A.L. Fetter and A.A. Svidzinsky, J.Phys.: Condens. Matter { \bf13}, R135
(2001);
A.J. Leggett, Rev. Mod. Phys. {\bf 73}, 307 (2001);
 K. Burnett, M. Edwards, and C.W. Clark, Phys. Today, {\bf 52}, 37 (1999);
 F. Dalfovo, S. Giorgini, L. Pitaevskii, and S. Stringari, Rev. Mod. Phys.
 {\bf 71}, 463 (1999);
S. Parkins, and D.F. Walls, Phys. Rep. {\bf 303}, 1 (1998).

\noindent
3.  L.P. Pitaevskii, Zh. Eksp. Teor. Fiz. {\bf 40}, 646 (1961) [Sov. Phys. JETP
 {
\bf 13}, 451 (1961)];
   E.P. Gross, Nuovo Cimento {\bf 20}, 454 (1961); J. Math. Phys. {\bf 4}, 195 (
1963).

\noindent
4. T.D. Lee and C.N. Yang, Phys. Rev. {\bf 105},1119 (1957);
T.D. Lee, K.W. Huang, and C.N. Yang, Phys. Rev. {\bf 106}, 1135 (1957).

\noindent
5. W. Ketterle, D.S. Durfee, and D.M. Stamper-Kurn, in  Proceedings of the  
International
 School of Physics ``Enrico Fermi", edited by M. Inquscio, S. Stringari,
 and
 C.E. Wieman (IOS Press, Amsterdam, 1999);
D.M. Stamper-Kurn, H.-J. Miesner, S. Inouye, M.R. Andrews, and W. Ketterle, 
Phys. Rev. Lett. {\bf81}, 500 (1998).

\noindent
6.  S. Inouye, M.R. Andrews, J. Stenger, H.J. Miesner, D.M. Stamper-Kurn, and
 W.
Ketterle, Nature {\bf 392}, 151 (1998);
P. Courteille, R.S. Freeland, D.J. Heinzen, F.A. van Abeelen, and B.J. Verhaar,
Phys. Rev. Lett. {\bf 81}, 69 (1998);
J.L. Roberts, N.R. Claussen, J.P. Burke, C.H. Greene, E.A. Cornell, and
C.E. Wieman, Phys. Rev. Lett. {\bf 81}, 5109 (1998);
D.M. Stamper-Kurn, and  W. Ketterle, Phys. Rev. Lett. {\bf 82}, 2422 (1999);
J.L. Roberts, N.R. Claussen, S.L. Cornish, E.A. Donley, E.A. Cornell, and
C.E. Wieman, Phys. Rev. Lett. {\bf 86}, 4211 (2001);
E.A. Donley, N.R. Claussen, S.L. Cornish, J.L. Roberts, E.A. Cornell, and
C.E. Wieman, Nature {\bf 412}, 295 (2001).

\noindent
7.  S.L. Cornish, N.R. Claussen, J.L. Roberts, E.A. Cornell, and  C.E. Wieman,
Phys. Rev. Lett. {\bf 85}, 1795 (2000).

\noindent
8.  J. Stenger, S. Inouye, M.R. Andrews, H.-J. Meisner, D.M. Stamper-Kurn, and W
. Ketterle, Phys. Rev. Lett. {\bf 82}, 2422 (1999).

\noindent
9. Y.E. Kim and A.L. Zubarev, Phys. Lett. A (In print); cond-mat/0212196.

\noindent
10.  J.H. Macek, Few-Body Systems {\bf 31}, 241 (2002).

\noindent
11. E. Timmermans, P. Tommasini, M,  Hussein, and A. Kerman,
Phys. Rep. {\bf 315}, 199 (1999).

\noindent
12. V.A, Yurovsky, A. Ben-Reuven, P.S.  Julienne,  and C.J. Williams,
Phys. Rev. A{\bf 60}, R765 (1999).

\noindent
13. F.A. van Abeelen  and B.J.  Verhaar, Phys. Rev. Lett. {\bf 83},1550 (1999).

\noindent
14. E. Timmermans, P. Tommasini, and K. Huang, Phys. Rev. A {\bf 55}, 3645
 (1997).

\noindent
15. E. Braaten and A. Nieto, Phys. Rev. B {\bf 56}, 14745 (1997);
E. Braaten and J. Pearson, Phys. Rev. Lett. {\bf 81}, 4541 (1998).

\noindent
16. A. Fabrocini and A. Polls, Phys. Rev. A {\bf60}, 2319 (1999);
Phys. Rev. A {\bf 64}, 063610 (2001).

\noindent
17. A. Banerjee and M.P. Singh, Phys. Rev. A{\bf64}, 063604 (2001);
Phys. Rev. A {\bf 66}, 043609 (2002).

\noindent
18. L. Pitaevskii and S. Stringari, Phys. Rev. Lett. {\bf 81}, 4541 (1998).

\noindent
19. Y.E. Kim and A.L. Zubarev, Phys. Rev. A {\bf 67}, 015602 (2003).
 
\noindent
20. E.A. Burt, R.W. Ghrist, C.J. Myatt, M.J. Holland, E.A. Cornell, and
C.E. Wieman, Phys. Rev. Lett. {\bf 79}, 337 (1997).

\noindent
21. Y. Kogan, B.V. Svistunov, and G.V. Shlyapnikov, JETP Lett. {\bf 42}, 209 (1985).

\noindent
22. D. Anderson, A. Bondeson, and M. Lisak, Phys. Lett. A {\bf 67}, 331 (1978).

\noindent
23. A. Bondeson, M. Lisak, and D. Anderson, Physica Scripta {\bf 20}, 479 (1979).

\noindent
24. S. Chavez Cerda, S.B. Cavalcanti, and J.M. Hickmann, Eur. J. Phys. D {\bf 1}, 313 (1998).

\noindent
25. W. Lenz, Z. Phys. {\bf 56}, 778 (1929).

\noindent
26. T.T. Wu, Phys. Rev. {\bf
115}, 1390 (1959).

\noindent
27. E.
Braaten, H.-W. Hammer, and T. Mehen, Phys. Rev. Lett. {\bf 88}, 040401 (2002)
and references therein.

\noindent
28.  S. Giorgini, J. Boronat, and J. Casulleras, Phys. Rev. A{\bf 60}, 5129 
(1999).

\noindent
29. G. Baum and C. Pethick, Phys. Rev. Lett. {\bf 76}, 6 (1006).

\noindent
30. E.H. Lieb and W. Liniger, Phys. Rev. {\bf 130}, 1605 (1963).

\noindent
31. D.M. Gangardt and G.V. Shlyapnikov, Phys. Rev. Lett. {\bf 90}, 010401 (2003).

\end{document}